\documentstyle[sprocl,epsfig]{article}

 5

 2
 3
 2

\def\nn{\nonumber\\}

\def\d{\partial}

\def\/{\hfill\break}
\def\be{\begin{equation}}
\def\ee{\end{equation}}
\def\ba{\begin{eqnarray}}
\def\ea{\end{eqnarray}}
\def\half{{1\over2}}

\def\exv#1{\left\langle #1\right\rangle}    
\def\mop#1{\mathop{\rm #1}\nolimits}        
\def\Tr{\mop{Tr}}
\def\ptlead{\leaders\hbox to 0.5em          
   {\hss .\hss}\hss}                        

\def\delv{\rlap{\lower-.5ex\hbox{--}}{\delta}}

\def\ket#1{\left| #1 \right\rangle}

\def\hphi{{\hat\Phi}}

\def\hh{{\hat H}}
\def\ho{{\hat O}}

\def\cO#1{{\cal O}\left(#1\right)}

\def\q{{\bf q}}
\def\p{{\bf p}}
\def\k{{\bf k}}

\def\fs{&&\hskip -0.6cm plus 0.1cm minus 0.1cm}

\def\DD{{\cal D}}
\def\Im{\mop{Im}}

\def\hob{\hbar\beta\omega}

\def\TUV{{\Tr}_{UV}}

\begin{document}
\title{ 
{\normalsize\mbox{ }\\\vspace{-2.3cm}
\hfill \parbox{30mm}{DESY 97-148\\\tt hep-ph/9708229}}\\[10mm]
CLASSICAL LIMIT IN SCALAR QFT AT HIGH TEMPERATURE}
\author{ANTAL JAKOV\'AC}
\address{Deutsches Elektronen-Synchrotron DESY\\D-22603 Hamburg, Germany}

\maketitle\abstracts{ It is shown that in a scalar quantum field
  theory at high temperatures we can compute even time dependent
  observables from an effective theory, which can be interpreted as a
  (nonlocal) classical statistical field theory. We examine the
  retarded Greens function in the local approximation, and define the
  classical self energy. Its imaginary part agrees with the quantum
  case in the leading order.}

\section{Introduction}

The complete description of a high temperature plasma concerns also
the computation of dynamical quantities (damping rates, topological
transitions etc.). Since there does not exist a direct nonperturbative
method to perform these calculations, it is worth to concentrate on
the most relevant degrees of freedom of the system and to use an
effective theory for them. At high temperatures the most relevant long
wavelength modes are highly excited, thus behave classically.
Therefore a good candidate for the effective theory in this regime is
the classical theory\cite{realtimesim}.

The classical theory, however, is plagued by UV divergences. In the
case of static observables the dimensional reduction program
\cite{dimred} solves this problem. There the classical model is
interpreted as a bare theory such that we get finite results for the
physical observables. The bare parameters can be calculated by
eliminating the UV modes of the system. The generalization to time
dependent quantities is, however, not straightforward. A natural
conjecture (c.f. Ref.\cite{RubShap}) is to use
$H_{eff}=\int[\half\pi^2+{\cal L}_{DR}(\phi)]$ as the effective
Hamiltonian where ${\cal L}_{DR}$ effective Lagrangian comes from the
dimensional reduction. Using this conjecture it has been
shown\cite{AartSmit} that the Greens functions of the scalar theory
are finite up to two loop order.

To better establish this idea we have to examine the properties of the
classical limit in hot quantum field theories. Firstly we have to find
some dimensionless quantity, which could govern this limit and
contains $\hbar$. In the free case it is $\hob_\k$, so we may try to
use this combination also in the perturbation theory. Then we can work
in $\hbar=1$ units. Secondly the time evolution of the classical
fields is deterministic, they are functions of the time and the
initial conditions. To arrive to a classical description we have to
integrate over the initial conditions, in the quantum language the
initial states.

\section{Effective theory}

Putting together this two requirements we will rewrite the thermal
average\cite{BuchJak}
\begin{equation}
  \exv{{\rm T}\,\hphi(x_1)\dots \hphi(x_1)}=\frac1Z\,\Tr\left(
  e^{-\beta\hh} {\rm T}\,\hphi(x_1)\dots \hphi(x_1)\right).
\end{equation}
As initial states we use coherent states $\ket\eta$, the
eigenstates of the annihilation operator at $t=t_0$. We choose a
separation scale $\Lambda$ ($m\ll\Lambda\ll T$). In the sense of the
Introduction the small parameter $\beta\Lambda$ stands for the
$\hbar$. We introduce the notations
\begin{eqnarray}
  \fs \TUV(\ho)=\int_{|\k|>\Lambda}\DD\eta\DD\eta^*
  \exv{\eta|\ho|\eta} \nn
  \fs \phi_\k=\frac1{\sqrt{2\omega_\k}}\biggl(\eta^*_{-\k}\,+\,
  \eta_\k \biggr) \qquad\qquad \pi_\k=i\, \frac{\sqrt{\omega_\k}}2
  \biggl(\eta^*_{-\k}\,-\, \eta_\k \biggr).
\end{eqnarray}
Then we can write
\begin{equation}
  \label{eff_mod_2}
  \exv{\ho}=\frac1Z\, \int_{|\k|<\Lambda}\!\!\!\DD\phi\DD\pi\,
  e^{-\beta H_{eff}}\, O_{eff},
\end{equation}
where
\begin{eqnarray}
\label{nota}
  \fs e^{-\beta H_{eff}}=\TUV e^{-\beta\hh}\nn
  \fs O_{eff}=\frac {\TUV e^{-\half \beta\hh}\ho\, e^{-\half
  \beta\hh}}{\TUV e^{-\beta\hh}}.
\end{eqnarray}
That is we have reduced the 4D path integration into two 3D path
integrations over a phase space (''dimensional reduction''), and we
can identify the effective action and the effective operator in this
reduced effective model.

The integration over the UV degrees of freedom is, as usual, supposed
to be perturbative. The omission of the IR initial conditions mean a
slight modification in the usual perturbation theory. For the scalar
field theory we find after some algebra
\begin{equation}
  \TUV e^{-\half \beta\hh}\ho\, e^{-\half \beta\hh}= e^{-i\int_c\!
    H_I(\frac \delta{i\delta j})}\,O\,(\frac \delta{i\delta j})\,
  e^{-\half\int_c jGj}\, e^{i\int_c j\tilde\phi_0}\biggr|_{j=0},
\end{equation}
where the integration contour is symmetric ($c:\, i/2\beta\to0\to
t_f\to 0 \to - i/2\beta$, $t_f$ is larger than any explicit time
arguments), $H_I$ is the interaction part of the Hamiltonian, changes sign
on the anti-time ordered part of the contour, and
\begin{eqnarray}
  \fs \tilde\phi_0(t,\k)=\Theta(\Lambda>|\k|)\,
  e^{-\half\beta\omega_\k}\left( \phi_\k\, \cos\omega_\k t\,+\,
  \pi_\k\, \frac{\sin\omega_\k t}{\omega_\k}\right)\nn
  \fs G(t_1,t_2,\k)=\frac1{2\omega_\k}\left[
  \Theta_c(t_1>t_2)\,e^{-i\omega_k(t_1-t_2)} \,+\,
  \Theta_c(t_2>t_1)\,e^{-i\omega_k(t_2-t_1)}\right]\,+\nn
  \fs \qquad\qquad\qquad\,+\, \Theta(|\k|>\Lambda) \,\frac{n(\omega_\k)}
  {\omega_\k} \,\cos\omega_\k (t_1-t_2),
\end{eqnarray}
($n(\omega)=[e^{\beta\omega}-1]^{-1}$). The perturbation theory for
the evaluation of the UV traces is like ordinary real time
perturbation theory, but in a background field $\tilde\phi_0$ which
follows a free time evolution, and with the propagator $G$, which
forbids the propagation of the IR thermal modes.

\section{IR and classical (local) approximation}

Similar to the $\hbar$ expansion in quantum mechanics we can now
expand with respect to $\beta\Lambda$. We concentrate on the leading
''classical'' order. We can omit the $e^{-\half\beta\omega_\k}$ factor
in the background field, its imaginary time evolution can also be
neglected (because $e^{\omega\beta}=1+\cO{\Lambda\beta}$). Another
consequence is the loss of quantum coherence, i.e. we can write for
the UV thermal expectation values
\[ \exv{{\rm T}\hphi(x_1)\dots\hphi(x_n)}_{UV}\to
\exv{\hphi(x_1)}_{UV} \dots \exv{\hphi(x_n)}_{UV}. \] 
This is because the IR dominant part of the complete propagator comes
from the effective theory, therefore in the computation of the UV
thermal expectation values all IR propagators can be cut (which does
not lead to vacuum diagrams). The complete proof will be published
elsewhere\cite{BuchJak}. Therefore the effective operator of the time
ordered n-point function can be substituted by $O_{eff}\to
\Phi(x_1)\dots\Phi(x_n)$, where $\Phi(x)$ is the effective one point
function (c.f. eq.~\ref{nota}).

Now we can use perturbation theory to determine $H_{eff}$ and
$\Phi(x)$. In $H_{eff}$ there is no time dependence, therefore
$t_f=0$, i.e., the Matsubara contour can be taken. The background is
the integration variable $\phi$, the conjugate momentum dependence
remains free. The rest of the computations is equivalent to
dimensional reduction, so finally we arrive at a form which agrees
with the conjecture mentioned in the Introduction.

For $\Phi(x)$ we can derive a gap equation, which can be written in a
differential equation form. The initial conditions
$\Phi(\k,t=0)=\phi_\k$,\ $\dot\Phi(\k,t=0)=\pi_\k$, and, symbolically,
\begin{equation}
  (\d_t^2+\omega_\k^2)\,\Phi= -\frac\lambda6\,\Phi\,-\, \Phi
\parbox[c]{0.9truecm}{\hfill
\setlength{\unitlength}{0.0125in}
\begin{picture}(20,20)(-10,-10) 
\put(0,0){\circle{20}}
\put(-11,0){\circle*{2}}
\end{picture}
\hfill}\,-\, \Phi
\parbox[c]{0.9truecm}{\hfill
\setlength{\unitlength}{0.0125in}
\begin{picture}(20,20)(-10,-10) 
\put(0,0){\circle{20}}
\put(-11,0){\circle*{2}}
\put(11,0){\circle*{2}}
\end{picture}
\hfill}\Phi^2.
\end{equation}
The first correction modifies the mass term to $m(T,\Lambda)$ in the
same way as in the effective Hamiltonian. The second term, however, is
nonlocal, we can denote it as $\lambda(x-x')/6$. We can define its
time-local part $\lambda_{loc}=\lambda(t=t')$, it is time independent,
and it is the same as the coupling of the effective Hamiltonian. The
nonlocal part after partial integration, with
$N(\omega)=\Theta(\omega-\Lambda) n(\omega)$
\begin{eqnarray}
  \lambda_{nonl}=\fs \int\!\frac{d^3\q}{(2\pi)^3}\,\frac1 {\omega_\q
    \omega_{\k-\q}} \Biggl( \frac{ 1 +N(\omega_\q) +
    N(\omega_{\k-\q})}
  {\omega_\q+\omega_{\k-\q}}\,\cos(\omega_\q+\omega_{\k-\q})
  (t-t')\,+\nn 
  \fs+\, \frac{ N(\omega_{\k-\q}) -N(\omega_\q)}
  {\omega_\q- \omega_{\k-\q}}\, \cos(\omega_\q-\omega_{\k-\q}) (t-t')
  \Biggr) \frac\d{\d t'}\,+\, \cO{\Lambda\beta}.
\end{eqnarray}

The local time evolution is determined by the effective Hamiltonian,
just as in the classical statistical mechanics.  Therefore a local
approximation provides at the same time the classical limit. The
quantum effects, as we could see, induce nonlocalities in the time
evolution.

We can also reintroduce $\hbar$ by substituting
\[ \beta\to\hbar\beta \qquad\qquad \lambda\to\hbar\lambda. \]
The neglected order is $\cO{\Lambda\beta\hbar}\sim\cO{\hbar}$, are the
terms which are suppressed by some power of $\hbar$. There are other
quantum corrections, however: some of them are powers of $1/\hbar$
(e.g. the one loop mass correction), others are just logarithmically
suppressed (as the nonlocal contribution to the damping rate
$\sim\sqrt\hbar\ln\hbar$). The IR effective theory contains both
corrections, the local (classical) approximation only the dominant
ones.

\section{The retarded Greens function}

Let us illustrate this with the calculation of the classical retarded
Greens function. The details can be found in
Refs\cite{BuchJak,BuchJak1}. We use the linear response theory: $H\to
H+\int j\phi$, which gives a current dependent solution $\Phi(x;j)$.
The classical retarded Greens function is
\begin{equation}
  D_R^{cl}(x-x')=\frac1Z\int\DD\phi\DD\pi\,e^{-\beta H}\,
  \frac{\delta\Phi(x;j)}{\delta j(x')}\biggr|_{j=0}.
\end{equation}
We can define the self energy from the Schwinger-Dyson equation
\begin{equation}
  D_R^{cl}(x-x')=D_R^{(0)}(x-x')\,+\, \int\!d^yd^4y'\,
  D_R^{(0)}(x-y)\, \Pi^{cl}(y-y')\,D_R^{cl}(y'-x'),
\end{equation}
where $D_R^{(0)}(\k,t)=-\Theta(t)\sin\omega_\k t/\omega_\k$. Its
imaginary part
\begin{eqnarray}
  \Im \Pi^{cl}(\omega,\p)\fs= \frac\pi2\, \lambda^2T^2\,\int
  \!\prod\limits_{i=1}^3 \frac{d^3\p_i}{(2\pi)^3\,2\omega_{i}}\,
  (2\pi)^3\delta(\Sigma\p_i-\p) \,\frac1{\omega_1\omega_2\omega_3}\nn
  \fs\qquad [\omega_1\,\delta(\omega-\omega_1-\omega_2-\omega_3)\,+\,
  \omega\, \delta(\omega+\omega_1-\omega_2-\omega_3)].
\end{eqnarray}
The imaginary part of the quantum self energy agrees with this result
at the leading order, if the effective parameters are used. The first
correction is a relative
$\cO{\sqrt{\lambda}\ln\lambda}=\cO{\sqrt{\hbar}\ln\hbar}$, which can
be reproduced from the nonlocal time evolution described earlier. On
the other hand the classical result can be thought of as the leading
term in the high temperature expansion of the quantum result. The
corrections then come from the subleading terms in the high
temperature expansion.

The quantum correction, being relatively suppressed as
$\sqrt{\lambda}\ln\lambda$, overwhelms the first classical correction
term. That limits the use of the classical theory: we have to keep
only the leading (nontrivial) order, and may average with respect to
the free Hamiltonian.

\section{Summary}

We have constructed an effective theory for computing time dependent
observables in scalar field theory. The form of the effective
Hamiltonian (after IR approximation) confirms the conjecture described
in the Introduction; the time evolution is, however, nonlocal. The
local approximation provides the usual classical limit, where the time
evolution is generated by the effective Hamiltonian. We have
calculated the classical retarded Greens function and defined the
classical self energy. Its imaginary part also agrees off-shell with
the corresponding quantum result. The first correction, however,
contains $\ln\hbar$, it is a genuine quantum effect. Hence, the use of
the classical theory beyond leading order does not appear meaningful.

\end{document}